\documentclass[11pt]{article}

\usepackage{epsfig,subfigure}
\usepackage{amssymb}
\usepackage{color}
\usepackage{hyperref}
\usepackage[cmex10,fleqn]{amsmath}

\usepackage{graphicx}
\usepackage{color}

\usepackage{amssymb}
\usepackage{amsmath,amsfonts,amssymb}

\usepackage{verbatim}
\usepackage{stfloats}
\usepackage[bookmarks=false]{}

\newtheorem{theorem}{Theorem}

\newtheorem{definition}{Definition}

\begin{document}
\date{}
\title{GDoF of the  MISO BC: Bridging the Gap between Finite Precision and Perfect CSIT}
\author{ \normalsize Arash Gholami Davoodi, Bofeng Yuan and Syed A. Jafar \\
{\small Center for Pervasive Communications and Computing (CPCC)}\\
{\small University of California Irvine, Irvine, CA 92697}\\
{\small \it Email: \{gholamid, bofengy, syed\}@uci.edu}
}
\maketitle

\begin{abstract}
For the $K=2$ user MISO BC, i.e., the wireless broadcast channel where a transmitter  equipped with $K=2$ antennas sends independent messages to $K=2$ receivers each of which is equipped with a single antenna,  the sum generalized degrees of freedom (GDoF) are characterized for arbitrary channel strength and channel uncertainty levels for each of the channel coefficients. The result is extended to $K>2$ users under  additional restrictions which include the assumption of symmetry. 
\end{abstract}
\newpage
\section{Introduction}
As the first steps in the path towards progressively refined capacity approximations, degrees of freedom (DoF) and generalized degrees of freedom (GDoF) studies of wireless networks have turned out to be surprisingly useful. By exposing large gaps where they exist in our understanding of the capacity limits, these studies have been the catalysts for numerous discoveries over the past decade \cite{Jafar_FnT}. Some of the most interesting unresolved questions brought to light by recent DoF and GDoF studies have to do with channel uncertainty and the diversity of channel strengths. Consider the wireless network with $K$ transmitters and $K$ receivers, which could represent the $K$ user interference channel, the $K\times K$ $X$ channel, or the $K$ user MISO BC, i.e., the broadcast channel formed by allowing full cooperation among the transmitters in a $K$ user interference channel. Consider, first the issue of channel uncertainty. If the channel state information at the transmitter(s) (CSIT) is perfect then the $K$ user interference channel has $K/2$ DoF, the $K\times K$ $X$ channel has $K^2/(2K-1)$ DoF, and the MISO BC has $K$ DoF almost surely.\footnote{Channel state information at the receivers (CSIR) is assumed perfect throughout this work.} The optimal DoF are achieved by interference alignment for the interference and $X$ channel settings, and by transmit zero-forcing in the MISO BC.  However, if the CSIT is available only to finite precision,  then the MISO BC has only $1$ DoF, i.e., the DoF collapse as conjectured by Lapidoth et al. nearly a decade ago   \cite{Lapidoth_Shamai_Wigger_BC}. The conjecture was proved recently in \cite{Arash_Jafar_GC14}. Since the MISO BC contains within it the $K$ user interference and $X$ channels, the collapse of DoF under finite precision CSIT implies that neither zero-forcing nor interference alignment is robust enough to provide a DoF advantage under finite precision CSIT, i.e., the DoF collapse for the interference and $X$ channels as well. Now consider the diversity of channel strengths which is explored through the studies of generalized degrees of freedom (GDoF). If all cross channels are much weaker relative to the direct channels then the GDoF do not collapse even with finite precision CSIT, e.g., the collapse of GDoF is avoided in the interference channel simply by treating the weak interference as noise \cite{Geng_TIN_opt}. Since the $X$ and BC settings include the interference channel, the collapse of DoF is avoided there as well. If some cross-channels are strong while others are so weak that they can be ignored entirely, as in the topological interference management problem \cite{Jafar_TIM}, then even under finite precision CSIT, interference alignment plays a key role, albeit in a more robust form that does not depend on actual channel realizations. 

Through these  isolated and somewhat extreme data points, the DoF and GDoF studies have established that the capacity of wireless networks in the high SNR regime is quite sensitive, separately, to the level of channel uncertainty and relative channel strengths. To build upon this progress, here we initiate a study that 1) spans the space between the extremes studied so far, and 2) unifies the isolated elements of the picture. To venture between the extremes we allow a range of channel knowledge spanning from perfect to absent, and a range of  channel strengths  spanning  from  weak to  strong. To present a unified view, we study the combined impact of both channel uncertainty and channel strengths by simultaneously incorporating both into our system model.

Arguably the main hurdle in expanding GDoF studies thus far has  been the difficulty of obtaining good outer bounds. This is exemplified by the conjecture of Lapidoth et al. which remained unresolved for nearly a decade. DoF outer bounds under channel uncertainty have until recently been limited mostly to compound channel arguments \cite{Weingarten_Shamai_Kramer}. Compound channel arguments produce tight outer bounds in several settings of interest that have been successfully explored in prior work. For example, it is known that in order to maintain the full DoF (i.e., the same as with perfect CSIT), the channel estimation error should scale as $O(SNR^{-1})$ \cite{Caire_Jindal_Shamai, Jindal, Kobayashi_Caire_Jindal}. Compound channel arguments also produce tight outer bounds for various settings involving retrospective  \cite{Gou_Jafar} and blind interference alignment \cite{Jafar_corr}. However, outer bounds based on compound channel arguments are evidently not strong enough to bridge the gap between perfect CSIT and finite precision CSIT. For instance, although the collapse of DoF of the MISO BC was originally conjectured under the compound setting by Weingarten et al. in \cite{Weingarten_Shamai_Kramer}, this conjecture was settled in the negative by \cite{Gou_Jafar_Wang} and \cite{Maddah_Compound}. 

The  reason that a broader study now seems feasible, is because of a new approach based on combinatorial accounting of the size of Aligned Image Sets (in short, the AIS approach), that was introduced  in \cite{Arash_Jafar_GC14} to  settle the conjectured collapse of DoF under finite precision CSIT. Reference  \cite{Arash_Jafar_GC14} also showed that the AIS approach could be used to address \emph{partial} CSIT. Consider the $K=2$ user MISO BC. For this setting the DoF are characterized in \cite{Arash_Jafar_GC14} with channel knowledge ranging from perfect to absent. In particular, if the channel estimation error terms scale as SNR$^{-\beta}$, so that $\beta$ values between $0$ and $1$ capture the full range of  channel uncertainties, from  essentially no channel knowledge ($\beta=0$) to  perfect channel knowledge ($\beta=1$), then it is shown that this channel has $1+\beta$ DoF. However, the study in \cite{Arash_Jafar_GC14} ignores the \emph{diversity} of channel knowledge since all channels are assumed to have the same $\beta$ parameter. Moreover, since this study is limited to DoF, it also does not capture the diversity of channel strengths\footnote{In the DoF model, any non-zero channel is capable of carrying only 1 DoF regardless of its strength.}. 

On the other hand, our recent work in \cite{Arash_Jafar_GC15} expands the AIS approach to study the diversity of channel strengths.  Consider again the $K=2$ user MISO BC. The sum GDoF for this setting are characterized under arbitrary channel strength levels for each of the channel coefficients in \cite{Arash_Jafar_GC15}. However, the study in \cite{Arash_Jafar_GC15} is limited to the extreme setting of $\beta=0$ for all channel coefficients, i.e., it does not capture the range and diversity of channel uncertainty parameters.  

Given these recent indicators that the AIS approach can be applied to study partial channel knowledge or the diversity of channel strengths individually, this work takes the next natural step, by jointly studying partial channel knowledge and the diversity of channel strengths in the same channel model. As a  result for the $K=2$ user setting, the sum generalized degrees of freedom (GDoF) are characterized for \emph{arbitrary} channel strength and channel uncertainty levels for each of the channel coefficients. Extensions to $K>2$ users are obtained under  additional assumptions of symmetry. 
 The results are presented and discussed in Section \ref{sec:mainresult}.

%
%
%
\section{System Model} {\label{sec-sys}}
\subsection{The Channel}
Under the GDoF framework, the channel model for the $K$ user MISO BC is defined by the following input-output equations.
\begin{eqnarray}
Y_k(t)=\sum_{l=1}^{K}\sqrt{P^{\alpha_{kl}}}G_{kl}(t)X_l(t)+Z_k(t),~~ \forall k\in[K].
\end{eqnarray}
The channel uses are indexed by $t\in\mathbb{N}$, $X_l(t)$ is the symbol sent from transmit antenna $l$ subject to a  unit power constraint, $Y_k(t)$ is the symbol observed by Receiver $k$, $Z_k(t)$ is the zero mean unit variance additive white Gaussian noise (AWGN) at Receiver $k$, and $G_{kl}(t)$ are the channel fading coefficients between transmit antenna $l$ and Receiver $k$. $P$ is the nominal $SNR$ parameter that is allowed to approach infinity. The channel strengths are represented in $\alpha_{kl}$ parameters. 
%

\subsection{Bounded Density Assumption}
An important definition for this work is the notion of a ``bounded density" assumption.
\begin{definition}[Bounded Density] A set of random variables, $\mathcal{A}$, is said to satisfy the bounded density assumption if 
 there exists a finite positive constant $f_{\max}$, \label{def1}
\begin{eqnarray*}
	0<f_{\max}<\infty
\end{eqnarray*}
such that for all finite cardinality disjoint subsets $\mathcal{A}_1, \mathcal{ A}_2$ of $\mathcal{A}$, 
\begin{eqnarray*}
	\mathcal{ A}_1\subset \mathcal{A}, \mathcal{A}_2\subset \mathcal{A}, \mathcal{A}_1\cap\mathcal{A}_2=\phi, \mathcal|{A}_1|<\infty, \mathcal|\mathcal{ A}_2|<\infty
\end{eqnarray*}
the conditional probability density functions exist and are bounded as follows,
\begin{eqnarray*}
	\forall A_1, A_2, ~~f_{\mathcal{A}_1|\mathcal{A}_2}(A_1|A_2)&\leq&f_{\max}^{|\mathcal{A}_1|}.
\end{eqnarray*}
\end{definition}


%

\subsection{Partial CSIT}
Under partial CSIT, the channel coefficients may be represented as
\begin{eqnarray*}
G_{kl}(t)&=&\hat{G}_{kl}(t)+\sqrt{P^{-\beta_{kl}}}\tilde{G}_{kl}(t)
\end{eqnarray*}
where $\hat{G}_{kl}(t)$ are the  channel estimate terms and $\tilde{G}_{kl}(t)$ are the  estimation error terms. To avoid degenerate conditions, the ranges of values are bounded away from zero and infinity as follows, i.e., there exist constants $\Delta_1, \Delta_2$ such that $0<\Delta_1\leq 
|{G}_{kl}(t)|$, and $|\tilde{G}_{kl}(t)|,|\tilde{G}_{kl}(t)|<\Delta_2<\infty$. The channel variables $\hat{G}_{kl}(t), \tilde{G}_{kl}(t)$, $\forall k,l\in\{1,2\}, t\in\mathbb{N}$, are subject to the bounded density assumption with the difference that the actual realizations of $\hat{G}_{kl}(t)$ are revealed to the transmitter, but the realizations of $\tilde{G}_{kl}(t)$ are not available to the transmitter.
Note that under the partial CSIT model, the variance of the channel coefficients $G_{kl}(t)$  behaves as $\sim P^{-\beta_{kl}}$ and the peak of the probability density function  behaves as $\sim\sqrt{P^{\beta_{kl}}}$.  In order to span the full range of partial channel knowledge at the transmitters, the corresponding range of $\beta_{kl}$ parameters, assumed throughout this work, is 
\begin{eqnarray}
0\leq \beta_{kl}\leq\alpha_{kl}
\end{eqnarray}
Note that  $\beta_{kl}=0$ and $\beta_{kl}=\alpha_{kl}$ correspond to the two extremes where the channel knowledge is essentially absent and perfect, respectively.
\subsection{GDoF}
The definitions of achievable rates $R_i(P)$ and capacity region $\mathcal{C}(P)$ are standard. The GDoF region is defined as
\begin{eqnarray}
\mathcal{D}&=&\{(d_1,\cdots,d_K): \exists (R_1(P),\cdots, R_K(P))\in\mathcal{C}(P), \mbox{ s.t. } d_k=\lim_{P\rightarrow\infty}\frac{R_k(P)}{C_o(P)}, \forall k\in[K]\} \label {region}
\end{eqnarray}
where $C_o(P)$ is a reference capacity of an additive white Gaussian noise channel $Y=X+N$ with transmit power $P$ and unit variance additive white Gaussian noise. For real settings, $C_o(P)=1/2\log(P)+o(\log(P))$ and for complex settings  $C_o(P)=\log(P)+o(\log(P))$.


\section{Main Results}\label{sec:mainresult}
%

\subsection{$K=2$ Users}
The first result is for the $K=2$ user MISO BC, where we allow arbitrary channel strength parameters $\alpha_{kl}$ and channel uncertainty parameters $\beta_{kl}$ for each channel coefficient. The sum GDoF for this setting is characterized in the following theorem.
\begin{theorem} \label{Theorem2}
The sum GDoF value of the $2$-user MISO BC is
\begin{eqnarray}
\mathcal{D}_\Sigma&=&\min(D_1,D_2)
\end{eqnarray}
where
\begin{eqnarray}
 D_1&=&\max(\alpha_{11},\alpha_{12})+\max(\alpha_{21}-\alpha_{11}+\min(\beta_{11},\beta_{12}),\alpha_{22}-\alpha_{12}+\min(\beta_{11},\beta_{12}),0)\\
 D_2&=&\max(\alpha_{21},\alpha_{22})+\max(\alpha_{11}-\alpha_{21}+\min(\beta_{21},\beta_{22}),\alpha_{12}-\alpha_{22}+\min(\beta_{21},\beta_{22}),0)
\label{aad}
\end{eqnarray}
\end{theorem}
Several observations can be made from Theorem \ref{Theorem2}. 
\begin{enumerate}
\item {\bf Recovering Prior Results:}
Since the current setting is a generalization of the $K=2$ user settings considered in \cite{Arash_Jafar_GC14} and \cite{Arash_Jafar_GC15}, naturally the corresponding results from \cite{Arash_Jafar_GC14} and \cite{Arash_Jafar_GC15} can be recovered as special cases of Theorem \ref{Theorem2}. For example, setting $\alpha_{ij}=1, \beta_{ij}=\beta$ for all $i,j\in\{1,2\}$, recovers the sum DoF result of \cite{Arash_Jafar_GC14}, i.e., 
\begin{eqnarray}
\mathcal{D}_\Sigma&=&1+\beta
\end{eqnarray}
Setting $\beta_{ij}=0$ for all $i,j\in\{1,2\}$, and allowing arbitrary $\alpha_{ij}$ values recovers the sum GDoF result of \cite{Arash_Jafar_GC15}, i.e.,
\begin{eqnarray}
\mathcal{D}_\Sigma&=&\min(D_1,D_2)\\
 D_1&=&\max(\alpha_{11},\alpha_{12})+\max((\alpha_{21}-\alpha_{11})^+,(\alpha_{22}-\alpha_{12})^+)\\
 D_2&=&\max(\alpha_{21},\alpha_{22})+\max((\alpha_{11}-\alpha_{21})^+,(\alpha_{12}-\alpha_{22})^+)
\end{eqnarray}

\item {\bf Redundancy of Strongest CSIT:} The sum GDoF value depends only on $\min(\beta_{11},\beta_{12})$ and $\min(\beta_{21},\beta_{22})$, i.e., it does not depend on the strongest CSIT parameter associated with each receiver. While for $K=2$ we can equivalently state that the GDoF \emph{depend only on the weakest} CSIT parameter for each receiver, it is easy to see\footnote{For instance, consider the $K=3$ user MISO BC where we have $3$ antennas at the transmitter and $\alpha_{ij}=1$ for all $i,j\in\{1,2,3\}$. Suppose there is no CSIT for all the channel coefficients associated with the first transmit antenna ($\beta_{11}=\beta_{21}=\beta_{31}=0$) and perfect CSIT ($\beta_{ij}=\alpha_{ij}$ for all $i\in\{1,2,3\}, j\in\{2,3\}$) for the rest. If the sum GDoF were limited by the worst case, then this setting would be equivalent to the case where all $\beta_{ij}=0$, i.e., $\mathcal{D}_\Sigma=1$ according to \cite{Arash_Jafar_GC14}. But we know that $2$ DoF are achievable simply by ignoring the first antenna and the first user, reducing it to a $2$ user MISO BC with perfect CSIT. } that such an interpretation does not extend beyond $K=2$ users. We expect that the insight that potentially generalizes to  $K>2$ users is that the GDoF value does \emph{not depend on the strongest} CSIT parameter for each receiver. Intuitively, this is because the receivers, with their full channel knowledge, have the ability to normalize one of the channel coefficients so that it is essentially known to the transmitter. Evidently, such a normalization can only be done for the channel coefficient with the strongest CSIT parameter without affecting the CSIT levels of the remaining coefficients. For compact notation let us define
\begin{eqnarray}
\beta_1&\triangleq&\min(\beta_{11},\beta_{12})\\
\beta_2&\triangleq&\min(\beta_{21},\beta{22})
\end{eqnarray}
Note that  $D_1, D_2$ may be equivalently expressed as
\begin{eqnarray}
D_1&=&\max(\alpha_{11},\alpha_{12},\alpha_{21}+(\alpha_{12}-\alpha_{11})^++\beta_1,\alpha_{22}+(\alpha_{11}-\alpha_{12})^++\beta_1)\\
D_2&=&\max(\alpha_{22},\alpha_{21},\alpha_{12}+(\alpha_{21}-\alpha_{22})^++\beta_2,\alpha_{11}+(\alpha_{22}-\alpha_{21})^++\beta_2)
\end{eqnarray}

\item {\bf Optimality of Single User Transmission:}
From the sum GDoF we note that if and only if both of the following conditions are satisfied
\begin{eqnarray}
\alpha_{11}&\geq&\alpha_{21}+\beta_1\\
\alpha_{12}&\geq&\alpha_{22}+\beta_1
\end{eqnarray}
then it is optimal to serve only user $1$, and the sum GDoF value is $\max(\alpha_{11},\alpha_{12})$. Note that the value of $\beta_2$ is irrelevant here. In words, it is optimal to serve only User 1, if and only if each transmit antenna `prefers' User 1 to User 2 (i.e., has a stronger connection to User 1 than User 2) by at least $\beta_1$. The corresponding conditions for optimality of serving only user $2$ are obtained by switching the indices.   Note that this is the only setting where all the available CSIT is useless.

\item {\bf GDoF vs CSIT Budget:} Since Theorem \ref{Theorem2} simultaneously allows  arbitrary levels of CSIT and arbitrary channels strengths, it offers insights into the optimal allocation of CSIT resources as a function of given channel strengths, which may be arbitrary, to maximize the sum GDoF. The CSIT budget formulation depends on the relative costs of acquiring CSIT for each link, which may depend on the feedback mechanism employed. As a simple example, suppose the total CSIT budget is 
\begin{eqnarray}
\beta=\beta_{11}+\beta_{12}+\beta_{21}+\beta_{22}
\end{eqnarray}
Then, given the value of $\beta$, it should be optimally allocated among $\beta_{11}, \beta_{12}, \beta_{21}, \beta_{22}$, as a function of all the channel strength parameters $\alpha_{11}, \alpha_{12}, \alpha_{21}, \alpha_{22}$, in order to maximize the sum GDoF value $\mathcal{D}_\Sigma(\beta)$. This can be done easily based on Theorem \ref{Theorem2}. For example, consider again the setting where \emph{each transmit antenna prefers the same user}, i.e.,
\begin{eqnarray}
\alpha_{11}&\geq&\alpha_{21}\\
\alpha_{12}&\geq&\alpha_{22}
\end{eqnarray}
Further, without loss of generality, let us assume that
\begin{eqnarray}
\alpha_{11}+\alpha_{22}&\geq&\alpha_{21}+\alpha_{12}
\end{eqnarray}
Note that there is no loss of generality in this assumption because the transmit antennas can always be labeled in a way that this assumption is true. Then, based on Theorem \ref{Theorem2}, the sum GDoF with the optimal allocation of CSIT are shown in Figure \ref{fig:GDoF(beta)}.

\begin{figure}[t]
\includegraphics[width=5.4in]{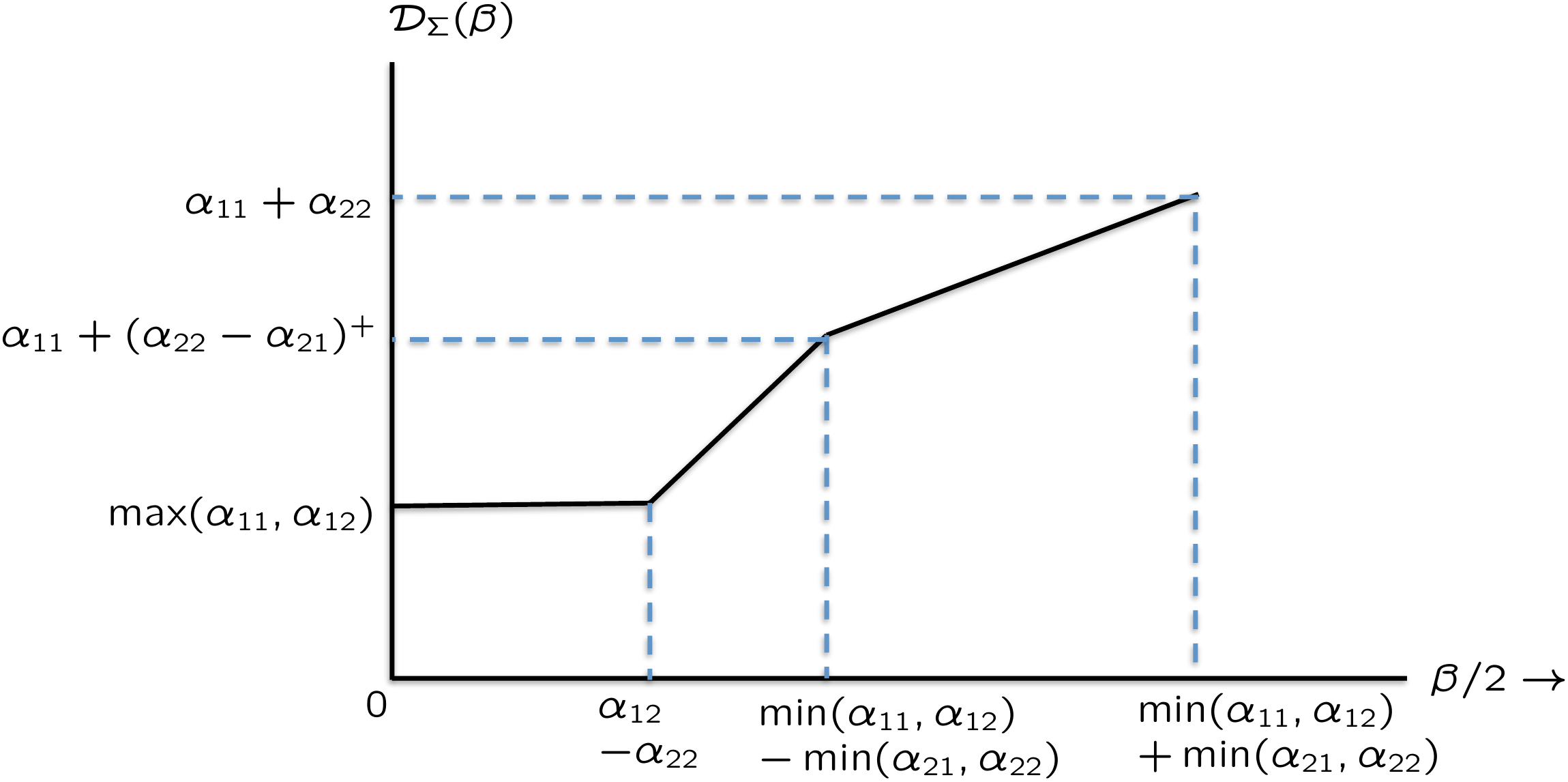}
\caption{Sum GDoF $\mathcal{D}_\Sigma(\beta)$ with optimal allocation of CSIT Budget $\beta$ when $\alpha_{11}\geq\alpha_{21}, \alpha_{12}\geq\alpha_{22}, \alpha_{11}+\alpha_{22}\geq\alpha_{21}+\alpha_{12}$.}\label{fig:GDoF(beta)}
\end{figure}

\item {\bf When each transmit antenna prefers a different user:} Consider the setting where each transmit antenna prefers a different user. Without loss of generality, suppose the first transmit antenna prefers User 1 and the second transmit antenna prefers User 2, i.e., 
\begin{eqnarray}
\alpha_{11}&>&\alpha_{21}\\
\alpha_{22}&>&\alpha_{12}
\end{eqnarray}
In this case, the sum GDoF value simplifies to
\begin{eqnarray}
\mathcal{D}_\Sigma&=&\min(\alpha_{22}+(\alpha_{11}-\alpha_{12})^++\beta_1, \alpha_{11}+(\alpha_{22}-\alpha_{21})^++\beta_2)
\end{eqnarray}
Note that in this case, increasing the CSIT budget $\beta$ always increases the sum GDoF under optimal CSIT allocation. In particular, if both `direct' channels are stronger than both 'cross' channels, i.e., 
\begin{eqnarray}
\min(\alpha_{11},\alpha_{22})&\geq&\max(\alpha_{12},\alpha_{21})
\end{eqnarray}
then
\begin{eqnarray}
\mathcal{D}_\Sigma&=&\alpha_{11}+\alpha_{22}-\max(\alpha_{12}-\beta_1, \alpha_{21}-\beta_2)
\end{eqnarray}
To see the sum GDoF with optimal allocation of CSIT resources, assume without loss of generality that $\alpha_{12}\geq\alpha_{21}$. The optimized sum GDoF $\mathcal{D}_\Sigma(\beta)$ in this case are shown in Figure \ref{fig:GDoF(beta)2}.
\begin{figure}[t]
\includegraphics[width=5.4in]{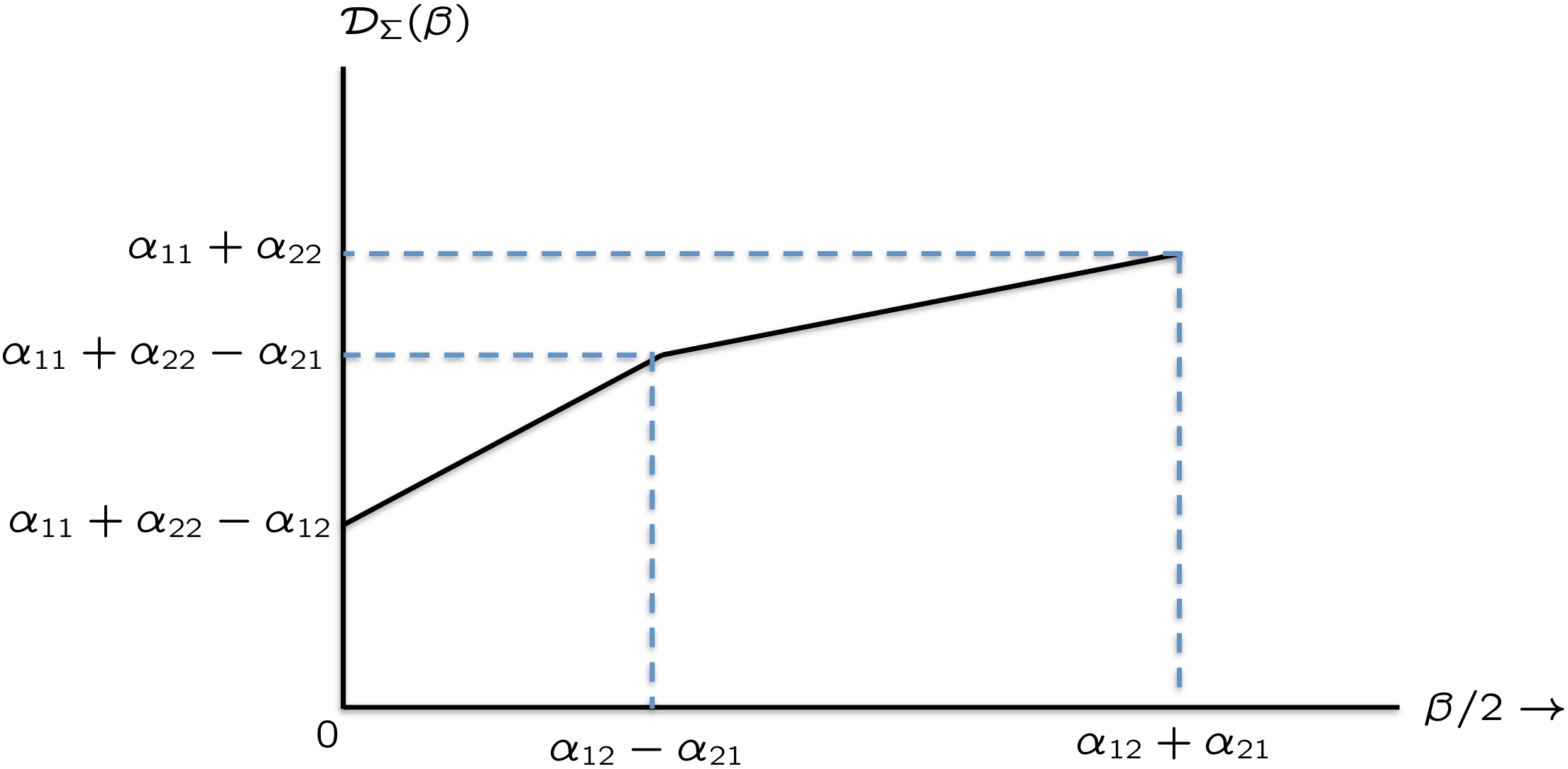}
\caption{Sum GDoF $\mathcal{D}_\Sigma(\beta)$ with optimal allocation of CSIT Budget $\beta$ when $\min(\alpha_{11},\alpha_{22})\geq\max(\alpha_{12},\alpha_{21}), \alpha_{12}\geq\alpha_{21}$.}\label{fig:GDoF(beta)2}
\end{figure}

\end{enumerate}
\subsection{Extension to $K$ Users}
The second result is an extension to the MISO BC with arbitrary number  of users ($K>2$), albeit under the following  restrictions which include assumptions of symmetry to limit the number of parameters. For all $k,l\in[K]$, we set
\begin{eqnarray}
\alpha_{kl}&=&\left\{
\begin{array}{ll}
\alpha, & k\neq l\\
1, &k=l.
\end{array}
\right., ~~
\alpha\in[0,1]\label{eq:cond1}\\
\beta_{kl}&=&\beta, ~~\beta\in[0,\alpha]\label{eq:cond2}
\end{eqnarray}
The GDoF characterization in this setting is presented in the following theorem.
\begin{theorem} \label{Theorem3}
The sum GDoF value of the $K$-user MISO BC that satisfies conditions (\ref{eq:cond1}) and (\ref{eq:cond2}) is
\begin{eqnarray}
\mathcal{D}_\Sigma&=&(\alpha-\beta)+K(1-(\alpha-\beta))
\end{eqnarray}
\end{theorem}
Recall that the DoF are obtained as a special case of GDoF, by setting $\alpha=1$. With this specialization, we note that Theorem \ref{Theorem3} shows that the $K$ user MISO BC has $1-\beta+K\beta$ DoF, matching the outer bound shown in \cite{Arash_Jafar_GC14}. This covers the extremes of perfect CSIT ($\beta=1$) where the DoF become equal to $K$ and finite precision CSIT ($\beta=0$) where the DoF collapse to $1$. It also shows that $\beta\geq 1$ is necessary to achieve the full $K$ DoF, thus matching the results of \cite{Jindal}. However, more significantly, it \emph{bridges} these divergent extremes by characterizing the DoF for all intermediate values of $\beta$ as well. 

The DoF value of $1-\beta+K\beta$ has a simple intuitive interpretation. Using  terminology analogous to  \cite{ADT_FnT}, the signal power levels  split into the bottom $\beta$ levels where CSIT is perfect and the remaining top $1-\beta$ levels where CSIT is only available to finite precision. This is because transmission in a direction orthogonal to estimated channel vector of undesired user (zero-forcing) with power up to $\sim P^\beta$ leaks no power above the noise floor at the undesired receiver. Due to essentially perfect zero-forcing, the bottom $\beta$ levels contribute $K\beta$ DoF. The top $1-\beta$ levels, which cannot be zero-forced, contribute the remaining $1-\beta$ DoF.\footnote{The achievability argument extends naturally to other settings. For example, it similarly follows that in the corresponding $K$ user interference channel the DoF value of $1-\beta+\frac{K}{2}\beta$ is  achievable.}

Beyond DoF, which implicitly assume all channels are equally strong,  by allowing $\alpha<1$, the GDoF setting allows us in this work to characterize the impact of different channel strengths (albeit restricted within assumptions of symmetry). Remarkably here we find that cross-channel strength parameters $\alpha$ and channel uncertainty parameters $\beta$ counter each other on equal terms, so that only their difference $(\alpha-\beta)$ matters. The sum GDoF value $(\alpha-\beta)+K(1-(\alpha-\beta))$ reflects essentially perfect CSIT over $1-(\alpha-\beta)$ dimensions which yield $K(1-(\alpha-\beta))$ GDoF through zero-forcing, while the remaining $(\alpha-\beta)$ dimensions cannot conceal interference and contribute only $(\alpha-\beta)$ GDoF. 

Finally, regarding the regime $\alpha>1$ which is not addressed in Theorem \ref{Theorem3}, we believe that this regime includes new challenges, both in terms of achievability and outer bounds, which go beyond the insights available so far.

\section{Proof of Theorem \ref{Theorem2}}
 The most interesting aspect of the proof  is the outer bound, for which we will generalize the Aligned Image Sets (AIS) argument of \cite{Arash_Jafar_GC14}. Since many of the details are repetitive we will focus primarily on the distinct aspects. Furthermore, we will present the proof only for the real setting here. Since the extension to complex settings follows along the lines of similar extensions in \cite{Arash_Jafar_GC14, Arash_Jafar_GC15} it does not bear repeating.
\subsection{Outer Bound}
For notational convenience, let us define
\begin{eqnarray}
\bar{P}&=&\sqrt{P}
\end{eqnarray}
The first step in the AIS approach is the transformation into a deterministic setting such that a GDoF outer bound on the deterministic setting is also a GDoF outer bound on the original setting. Since the derivation of the deterministic setting is identical to \cite{Arash_Jafar_GC14}, we directly present the deterministic model as follows.

\subsubsection{Deterministic Channel Model}
The deterministic channel model has inputs $\bar{X}_i(t) \in\mathbb{Z}$ and outputs $\bar{Y}_i(t) \in\mathbb{Z},~ \forall t\in\mathbb{N},  i \in \{1,2\}$, such that
\begin{eqnarray}
\bar{Y}_1(t)&=&\lceil \bar{P}^{\alpha_{11}-\max(\alpha_{11},\alpha_{12})} G_{11}(t)\bar{X}_1(t)\rceil+\lceil \bar{P}^{\alpha_{12}-\max(\alpha_{11},\alpha_{12})}G_{12}(t)\bar{X}_2(t)\rceil\\
\bar{Y}_2(t)&=&\lceil\bar{P}^{\alpha_{21}-\max(\alpha_{21},\alpha_{22})}G_{21}(t)\bar{X}_1(t)\rceil+\lceil \bar{P}^{\alpha_{22}-\max(\alpha_{21},\alpha_{22})}G_{22}(t)\bar{X}_2(t)\rceil~~~~~~~
\end{eqnarray}
and $\bar{X}_1(t)\in\{0,1,\cdots,\lceil \bar{P}^{\max(\alpha_{11},\alpha_{12})} \rceil\}$, $\bar{X}_2(t)\in\{0,1,\cdots,\lceil \bar{P}^{\max(\alpha_{21},\alpha_{22})} \rceil\}$.

\subsubsection{Functional Dependence and Aligned Image Sets}
Following directly along the AIS approach \cite{Arash_Jafar_GC14}, and omitting  $o(\log(P))$ and $o(n)$ terms that are inconsequential for GDoF, we have 
\begin{eqnarray}
n(R_1+R_2)&\leq& H(\bar{Y}_1^{[n]}|W_2,G^{[n]})+H(\bar{Y}_2^{[n]}|G^{[n]})-H(\bar{Y}_2^{[n]}|G^{[n]},W_2)\\
&\leq& n\max(\alpha_{21},\alpha_{22})\log(\bar{P})+H(\bar{Y}_1^{[n]}|W_2,G^{[n]})-H(\bar{Y}_2^{[n]}|G^{[n]},W_2)\label{eq:fd1}
\end{eqnarray}
As in \cite{Arash_Jafar_GC14}, for the outer bound there is no loss of generality in fixing $W_2$ as a constant, and assuming the following functional dependence
\begin{eqnarray}
(\bar{X}_1^{[n]},\bar{X}_2^{[n]})&=&f_1(\bar{Y}_1^{[n]},G_{11}^{[n]},G_{12}^{[n]})\\
\Rightarrow\bar{Y}_2^{[n]}&=&f_2(\bar{Y}_1^{[n]},G^{[n]})
\end{eqnarray}
The aligned image sets are defined as
\begin{align}
S_{\nu^{[n]}}(G^{[n]})&=\{\bar{Y}_1^{[n]}\mbox{ s. t. } f_2(\bar{Y}_1^{[n]},G^{[n]})=f_2(\nu^{[n]},G^{[n]})\}
\end{align}
i.e.,  $S_{\nu^{[n]}}(G^{[n]})$ is the set of distinct images --- one of which is $\nu^{[n]}$ --- cast at Receiver 1,  which correspond to the same image at Receiver 2. 

Following the AIS approach, the sum-rate bound in (\ref{eq:fd1}) leads to the following bound expressed in terms of the expected cardinality of the aligned image sets.
\begin{eqnarray}
n(R_1+R_2)&\leq&  n\max(\alpha_{21},\alpha_{22})\log(\bar{P})+\log\mbox{E}\left|S_{\nu^{[n]}}(G^{[n]})\right|\label{eq:Fano}
\end{eqnarray}

\subsubsection{Bounding the Probability that Images Align}
Given $G_{11}^{[n]}, G_{12}^{[n]}$, consider two distinct realizations of User 1's output sequence $\bar{Y}_1^{[n]}$, denoted as $\lambda^{[n]}$ and $\nu^{[n]}$, which are produced by the corresponding  realizations of the codeword $(X_1^{[n]}, X_2^{[n]})$ denoted by $(\lambda_1^{[n]},\lambda_2^{[n]})$ and $(\nu_1^{[n]}, \nu_2^{[n]})$, respectively. 
\begin{eqnarray}
\lambda(t)&=&\lfloor \bar{P}^{\alpha_{11}-\max(\alpha_{11},\alpha_{12})} G_{11}(t)\lambda_1(t)\rfloor+\lfloor \bar{P}^{\alpha_{12}-\max(\alpha_{11},\alpha_{12})}G_{12}(t)\lambda_2(t)\rfloor\label{eq:lambdanu1}\\
\nu(t)&=&\lfloor \bar{P}^{\alpha_{11}-\max(\alpha_{11},\alpha_{12})}G_{11}(t)\nu_1(t)\rfloor
+\lfloor \bar{P}^{\alpha_{12}-\max(\alpha_{11},\alpha_{12})}G_{12}(t)\nu_2(t)\rfloor\label{eq:lambdanu2}~~~~~~~
\end{eqnarray}
We wish to bound the probability that the images of these two codewords align at User 2, i.e., $\nu^{[n]}\in S_{\lambda^{[n]}}$. For simplicity, consider first the single channel use setting, $n=1$. For $\nu\in S_\lambda$ we must have,
\begin{eqnarray}
&&\lfloor\bar{P}^{\alpha_{21}-\max(\alpha_{21},\alpha_{22})}G_{21}\nu_1\rfloor+\lfloor \bar{P}^{\alpha_{22}-\max(\alpha_{21},\alpha_{22})}G_{22}\nu_2\rfloor\nonumber\\
&=&\lfloor \bar{P}^{\alpha_{21}-\max(\alpha_{21},\alpha_{22})}G_{21}\lambda_1\rfloor+\lfloor \bar{P}^{\alpha_{22}-\max(\alpha_{21},\alpha_{22})} G_{22}\lambda_2\rfloor\label{xe}
\end{eqnarray}
So for fixed value of $G_{22}$ the random variable $\bar{P}^{\alpha_{21}-\max(\alpha_{21},\alpha_{22})}G_{21}(\nu_1-\lambda_1)$ must take values within an interval of length no more than 4.  If $\nu_1\neq\lambda_1$, then $G_{21}$ must take values in an interval of length no more than $\frac{4}{\bar{P}^{\alpha_{21}-\max(\alpha_{21},\alpha_{22})}|\nu_1-\lambda_1|}$, the probability of which is no more than $\frac{4f_{\max}{\bar{P}}^{\beta_{21}}}{\bar{P}^{\alpha_{21}-\max(\alpha_{21},\alpha_{22})}|\nu_1-\lambda_1|}$. Similarly, for fixed value of $G_{21}$ the random variable $\bar{P}^{\alpha_{22}-\max(\alpha_{21},\alpha_{22})}G_{22}(\nu_2-\lambda_2)$ must take values within an interval of length no more than 4.   If $\nu_1=\lambda_1$ then, because $\nu \neq \lambda$, we must have $\nu_2\neq\lambda_2$, and the probability of alignment is similarly bounded by $\frac{4f_{\max}{\bar{P}}^{\beta_{22}}}{\bar{P}^{\alpha_{22}-\max(\alpha_{21},\alpha_{22})}|\nu_2-\lambda_2|}$. Thus, based on $(\ref{xe})$, either the probabilty of alignment is zero or we have, 
\begin{eqnarray}
\bar{P}^{\alpha_{21}-\max(\alpha_{21},\alpha_{22})}\Delta_1|\nu_1-\lambda_1|&\leq& \bar{P}^{\alpha_{22}-\max(\alpha_{21},\alpha_{22})} \Delta_2|\nu_2-\lambda_2|+2\\
 \bar{P}^{\alpha_{22}-\max(\alpha_{21},\alpha_{22})}\Delta_1|\nu_2-\lambda_2|&\leq&   \bar{P}^{\alpha_{21}-\max(\alpha_{21},\alpha_{22})}\Delta_2|\nu_1-\lambda_1|+2
\end{eqnarray}
Next we will bound the max of $ \bar{P}^{\alpha_{21}-\max(\alpha_{21},\alpha_{22})}|\nu_1-\lambda_1|$ and $ \bar{P}^{\alpha_{22}-\max(\alpha_{21},\alpha_{22})}|\nu_2-\lambda_2|$. From (\ref{eq:lambdanu1}) and (\ref{eq:lambdanu2}) we have
\begin{eqnarray}
|\lambda-\nu|&\leq&2+ \bar{P}^{\alpha_{11}-\max(\alpha_{11},\alpha_{12})}|G_{11}||\lambda_1-\nu_1|+ \bar{P}^{\alpha_{12}-\max(\alpha_{11},\alpha_{12})}|G_{12}||\lambda_2-\nu_2|~~~\\
&\leq&2+2\Delta_2 \max(\bar{P}^{\alpha_{11}-\max(\alpha_{11},\alpha_{12})}|\nu_1-\lambda_1|,\bar{P}^{\alpha_{12}-\max(\alpha_{11},\alpha_{12})} |\nu_2-\lambda_2|)~~~\\
&\leq&2+2\Delta_2 \max(\bar{P}^{\alpha_{21}-\max(\alpha_{21},\alpha_{12})}|\nu_1-\lambda_1|,\bar{P}^{\alpha_{22}-\max(\alpha_{11},\alpha_{12})} |\nu_2-\lambda_2|)~~~\nonumber\\
&&\times \bar{P}^{\max(\alpha_{11}-\alpha_{21},\alpha_{12}-\alpha_{22})}
\end{eqnarray}
so, if $|\lambda-\nu|>\frac{4\Delta_2\bar{P}^{\max(\alpha_{11}-\alpha_{21},\alpha_{12}-\alpha_{22})}}{\Delta_1}+2$, the probability of $\nu\in S_\lambda$ is no more than
\begin{eqnarray}
&&\frac{4\Delta_2f_{\max}{\bar{P}}^{\beta_{22}}}{\bar{P}^{\alpha_{22}-\max(\alpha_{21},\alpha_{22})}\Delta_2|\nu_2-\lambda_2|}\\
&\leq&\frac{4\Delta_2f_{\max}{\bar{P}}^{\beta_{22}}}{\max(\bar{P}^{\alpha_{21}-\max(\alpha_{21},\alpha_{22})}\Delta_1|\nu_1-\lambda_1|-2,\bar{P}^{\alpha_{22}-\max(\alpha_{21},\alpha_{22})}\Delta_2|\nu_2-\lambda_2|)}~~~~~\\
&\leq&\frac{4\Delta_2f_{\max}{\bar{P}}^{\beta_{22}}}{\Delta_{1}\max(\bar{P}^{\alpha_{21}-\max(\alpha_{21},\alpha_{22})}|\nu_1-\lambda_1|,\bar{P}^{\alpha_{22}-\max(\alpha_{21},\alpha_{22})}|\nu_2-\lambda_2|)-2}~~~~~\\
&\leq&\frac{4\Delta_2f_{\max}{\bar{P}}^{\beta_{22}}}{\Delta_1\frac{|\lambda-\nu|-2}{2\Delta_2\bar{P}^{\max(\alpha_{11}-\alpha_{21},\alpha_{12}-\alpha_{22})}}-2}\\
&\leq&\frac{8\frac{\Delta_2^2}{\Delta_1}f_{\max}{\bar{P}}^{\max(\alpha_{11}-\alpha_{21},\alpha_{12}-\alpha_{22})+\beta_{22}}}{|\lambda-\nu|-\frac{4\Delta_2\bar{P}^{\max(\alpha_{11}-\alpha_{21},\alpha_{12}-\alpha_{22})}}{\Delta_1}-2}
\end{eqnarray}
Define $\Delta=\frac{4\Delta_2\bar{P}^{\max(\alpha_{11}-\alpha_{21},\alpha_{12}-\alpha_{22})}}{\Delta_1}+2$, ($\Delta$ scales as $\bar{P}^{\max(\alpha_{11}-\alpha_{21},\alpha_{12}-\alpha_{22})}$). Now let us return to the case of general $n$, where we similarly have,
\begin{eqnarray}
\mathbb{P}(\lambda^{[n]}\in S_{\nu^{[n]}})&\leq&\prod_{t:|\lambda(t)-\nu(t)|\leq\Delta} 1\times \prod_{t:|\lambda(t)-\nu(t)|>\Delta}\frac{8\frac{\Delta_2^2}{\Delta_1}f_{\max}\bar{P}^{\max(\alpha_{11}-\alpha_{21},\alpha_{12}-\alpha_{22})+\beta_{22}}}{|\lambda(t)-\nu(t)|-\Delta}\nonumber
\end{eqnarray}
\subsubsection{Bounding the Expected Size of Aligned Image Sets.}
\begin{eqnarray}
\mbox{E}(|S_{\nu^{[n]}}|)&=&\sum_{\lambda^n\in\{\bar{Y_1}^{[n]}\}}\mathbb{P}\left(\lambda^n\in S_{\nu^{[n]}}\right)\nonumber\\
&\leq&\prod_{t=1}^n\left(\sum_{\lambda(t):|\lambda(t)-\nu(t)|\leq\Delta}1+\sum_{\lambda(t):|\lambda(t)-\nu(t)|>\Delta}\frac{8\frac{\Delta_2^2}{\Delta_1}f_{\max}\bar{P}^{\max(\alpha_{11}-\alpha_{21},\alpha_{12}-\alpha_{22})+\beta_{22}}}{|\lambda(t)-\nu(t)|-\Delta}\right)\nonumber\\
&\leq&\prod_{t=1}^n\left(2\Delta+1+{8\frac{\Delta_2^2}{\Delta_1}f_{\max}\bar{P}^{\max(\alpha_{11}-\alpha_{21},\alpha_{12}-\alpha_{22})+\beta_{22}}} \times 2(1+ \max(\alpha_{11},\alpha_{12})\log (1+2\Delta_2\bar{P}))\right)\nonumber\\
&\leq&(8\frac{\Delta_2^2}{\Delta_1}f_{\max})^n\bar{P}^{n(\max(\alpha_{11}-\alpha_{21},\alpha_{12}-\alpha_{22})+\beta_{22})^+}\times\left(\max(\alpha_{11},\alpha_{12})\log(\bar{P})+o(\log(\bar{P}))\right)^n\label{eq:aveSK}\nonumber
\end{eqnarray}
\subsubsection{The GDoF Bound}
Substituting back into (\ref{eq:Fano}) we have
\begin{eqnarray}
n(R_1+R_2)&\leq &n\max(\alpha_{21},\alpha_{22})\log(\bar{P})+\log\mbox{E}|S_{\nu^{[n]}}|\nonumber\\
&\leq &n\left(\max(\alpha_{21},\alpha_{22})+(\max(\alpha_{11}-\alpha_{21},\alpha_{12}-\alpha_{22})+\beta_{22})^+\right)\log(\bar{P})\nonumber
\end{eqnarray}
So that we obtain the GDoF bound
\begin{eqnarray}
d_1+d_2&\leq&\max(\alpha_{21},\alpha_{22})+\max(\alpha_{11}-\alpha_{21}+\beta_{22},\alpha_{12}-\alpha_{22}+\beta_{22},0)
\end{eqnarray}
By symmetry we also have the GDoF bounds,
\begin{eqnarray}
d_1+d_2&\leq&\max(\alpha_{21},\alpha_{22})+\max(\alpha_{11}-\alpha_{21}+\beta_{21},\alpha_{12}-\alpha_{22}+\beta_{21},0)\\
d_1+d_2&\leq&\max(\alpha_{11},\alpha_{12})+\max(\alpha_{21}-\alpha_{11}+\beta_{11},\alpha_{22}-\alpha_{12}+\beta_{11},0)\\
d_1+d_2&\leq&\max(\alpha_{11},\alpha_{12})+\max(\alpha_{21}-\alpha_{11}+\beta_{12},\alpha_{22}-\alpha_{12}+\beta_{12},0)
\end{eqnarray}
Note that $\min(M+\max(A,0),M+\max(B,0))=M+\max(\min(A,B),0)$, so, together these bounds give us $d_1+d_2\leq \min(D_1,D_2)$, completing the proof of the outer bound for Theorem \ref{Theorem2}.

\subsection {Achievability}
Since the GDoF depend only on the worst channel uncertainty of each receiver, i.e., $\min(\beta_{11},\beta_{12})$ for receiver $1$ or $\min(\beta_{21},\beta_{22})$ for receiver $2$, for the achievability proof, we can  assume without loss of generality that $\beta_{11},\beta_{12}$ are equal to $\beta_1$, and, $\beta_{21},\beta_{22}$ are equal to $\beta_2$. With this assumption we will prove that $\min(D_1,D_2)$ is achievable.
Without loss of generality, we ignore measure zero events such as channel rank-deficiencies. This is because the channels are generated according to bounded densities, so that the probability mass that can be placed in a space whose measure approaches zero, must also approach zero. 

Without loss of generality assume $\alpha_{11}$ as the maximum of $\alpha_{ij}$,  $\forall i,j \in \{1,2\}$. The achievability proof is presented separately for the three cases of $(\alpha_{21}>\alpha_{22}, \alpha_{11}-\alpha_{12}>\alpha_{21}-\alpha_{22})$; $(\alpha_{21}>\alpha_{22}, \alpha_{11}-\alpha_{12}\leq \alpha_{21}-\alpha_{22})$; and $(\alpha_{21}\leq\alpha_{22})$.
\begin {enumerate}
\item {$\alpha_{21}>\alpha_{22}, \alpha_{11}-\alpha_{12}>\alpha_{21}-\alpha_{22}$.}\\
We wish to achieve the sum-DoF value of $d_1 + d_2 =\min(\alpha_{11}+(\alpha_{22}-\alpha_{12}+\beta_1)^+,\alpha_{11}+\beta_2)$. As $ \alpha_{11}-\alpha_{12}>\alpha_{21}-\alpha_{22}$, the first antenna can transmit $\alpha_{21}-\alpha_{22}$ DoF using highest power level as it will be decoded at the receivers without any interference from the second antenna. So, decreasing both $\alpha_{11},\alpha_{21}$ by $\alpha_{21}-\alpha_{22}$, we have a new channel with channel coefficients $\alpha'_{11}=\alpha_{11}-\alpha_{21}+\alpha_{22},\alpha'_{12}=\alpha_{12},\alpha'_{21}=\alpha_{22},\alpha'_{22}=\alpha_{22}$ where, we need to achieve  the sum-DoF value of $d_1 + d_2 =\min(\alpha'_{11}+(\alpha'_{21}-\alpha'_{12}+\beta_1)^+,\alpha'_{11}+\beta_2)$ through the tuple $d_1 = \alpha'_{11}, d_2 =m$, where $m=\min((\alpha'_{21}-\alpha'_{12}+\beta_1)^+,\beta_2)$. The case $m=0$ is obviously achieved, So, lets consider the case where $m>0$ i.e. $\alpha'_{21}-\alpha'_{12}+\beta_1>0$. To achieve $\alpha'_{11}+m$ DoF, let us split User 1's message
as $W_1 = (W_c,W_{1z},W_{1p})$ and User 2's message as $W_{2} = W_{2z}$,
where $W_{1z},W_{1p}$ act as private sub-messages to be decoded
only by user 1, $W_{2z}$ acts as a private sub-message to be
decoded only by User 2, while $W_c$ acts as a common submessage
that can be decoded by both users. $W_c$, $W_{1z}$, $W_{2z}$ and $W_{1p}$
carry $\alpha'_{21}-m,m,m,\alpha'_{11}-\alpha'_{21}$ DoF respectively. Messages $W_c,W_{1z},W_{2z},W_{1p}$ are encoded into
independent Gaussian codebooks $X_c,X_{1z},X_{2z},X_{1p}$, with unit powers, producing the transmitted symbols as follows.
\begin{align}
\left[\begin{array}{c}
X_1\\
X_2
\end{array}
\right]=&~~c_o{\bf V}_cX_c+c_o\sqrt{P^{-\alpha'_{21}}}{\bf V}_{1p}X_{1p}+c_o\sqrt{P^{m-\alpha'_{21}}}{\bf V}_{1z} X_{1z}+c_o{\bf V}_{2z} X_{2z}
\end{align}
Here ${\bf V}_{c}$, ${\bf V}_{1p}$, ${\bf V}_{1z}$, ${\bf V}_{2z}$ are vectors as follows
\begin{eqnarray}
{\bf V}_{c}&=&\left[\begin{array}{c}
1\\
0
\end{array}
\right]\\
{\bf V}_{1p}&=&\left[\begin{array}{c}
1\\
0
\end{array}
\right]\\
{\bf V}_{1z}&=&\left[\begin{array}{c}
\hat{G}_{22}\\
-\hat{G}_{21}
\end{array}
\right]\\
{\bf V}_{2z}&=&\left[\begin{array}{c}
\hat{G}_{12}\sqrt{P^{m+\alpha'_{12}-\alpha'_{21}-\alpha'_{11}}}\\
-\hat{G}_{11}\sqrt{P^{m-\alpha'_{21}}}
\end{array}
\right]
\end{eqnarray}
In words, ${\bf V}_{1z}$ is a unit vector orthogonal to the estimated channel vector of User $2$, and ${\bf V}_{2z}$ is a unit vector orthogonal to the estimated channel vector of User 1. Thus,  $X_{1z},X_{2z}$ are  zero-forced to the estimated channels of the undesired users. $c_o$ is a scaling factor, $O(1)$ in $P$, chosen to ensure that the transmit power constraint is satisfied. 
The  signal seen at Receiver 1 is,
\begin{eqnarray}
{Y}_1&=&\left[\begin{array}{cc}\bar{P}^{\alpha'_{11}}\hat{G}_{11}&\bar{P}^{\alpha'_{12}}\hat{G}_{12}\end{array}\right]\left[\begin{array}{c}
X_1\\
X_2
\end{array}
\right]+
\left[\begin{array}{cc}\bar{P}^{\alpha'_{11}-\beta_1}\tilde{G}_{11}&\bar{P}^{\alpha'_{12}-\beta_1}\tilde{G}_{12}\end{array}\right]\left[\begin{array}{c}
X_1\\
X_2
\end{array}
\right]+Z_1\nonumber\\
&=&\bar{P}^{\alpha'_{11}}c_1 X_c+\bar{P}^{\alpha'_{11}-\alpha'_{21}}c_2X_{1p}+\bar{P}^{\alpha'_{11}+m-\alpha'_{21}}c_3X_{1z}+\bar{P}^{m+\alpha'_{12}-\beta_1-\alpha'_{21}}c_4X_{2z}+Z_1\nonumber
\end{eqnarray}
where the $c_i$ are non-zero and bounded, i.e., $O(1)$  functions of $P$. Note that, $m+\alpha'_{12}-\beta_1-\alpha'_{21}\leq 0$. User 1 first decodes $X_c$ while treating all other signals as white noise. This is possible because  $X_c$ is received with power $\sim {P}^{\alpha'_{11}}$, the effective noise has power $\sim P^{\alpha'_{11}+m-\alpha'_{21}}$, and $X_c$ carries $\alpha'_{21}-m$ DoF. After decoding $X_c$, the receiver subtracts its contribution from its received signal and then proceeds to decode $X_{1z}$ while treating remaining signals as noise. This is possible since $X_{1z}$ is received with power $\sim {P}^{\alpha'_{11}+m-\alpha'_{21}}$,  the effective noise has power $\sim {P}^{\alpha'_{11}-\alpha'_{21}}$, and $X_{1z}$ carries $m$ DoF.   After decoding $X_{1z}$, the receiver subtracts its contribution from its received signal and then proceeds to decode $X_{1p}$ while treating remaining signals as noise. Since $X_{1p}$ is received with power $\sim {P}^{\alpha'_{11}-\alpha'_{21}}$, 
the remaining signals and noise are received with only  $O(1)$ power, and $X_{1p}$ carries $\alpha'_{11}-\alpha'_{21}$ DoF, this decoding is successful as well.  The  signal seen at Receiver 2 is,
\begin{eqnarray}
{Y}_2&=&\left[\begin{array}{cc}\bar{P}^{\alpha'_{21}}\hat{G}_{21}&\bar{P}^{\alpha'_{22}}\hat{G}_{22}\end{array}\right]\left[\begin{array}{c}
X_1\\
X_2
\end{array}
\right]+
\left[\begin{array}{cc}\bar{P}^{\alpha'_{21}-\beta_2}\tilde{G}_{21}&\bar{P}^{\alpha'_{22}-\beta_2}\tilde{G}_{22}\end{array}\right]\left[\begin{array}{c}
X_1\\
X_2
\end{array}
\right]+Z_2\nonumber\\
&=&\bar{P}^{\alpha'_{21}}e_1 X_c+e_2X_{1p}+\bar{P}^{m-\beta_2}e_3X_{1z}+\bar{P}^{m}e_4X_{2z}+Z_1\nonumber
\end{eqnarray}
where the $e_i$ are non-zero and bounded, i.e., $O(1)$  functions of $P$. Note that, $m-\beta_2\leq 0$. User 1 first decodes $X_c$ while treating all other signals as white noise. This is possible because  $X_c$ is received with power $\sim {P}^{\alpha'_{21}}$, the effective noise has power $\sim P^{m}$, and $X_c$ carries $\alpha'_{21}-m$ DoF. After decoding $X_c$, the receiver subtracts its contribution from its received signal,  and then decodes $X_{2z}$. Since $X_{2z}$ is received with power $\sim {P}^{m}$,  the remaining signals and noise are received with only  $O(1)$ power, and as $X_{2z}$ carries $m$ DoF, this decoding is successful as well.

\item {$\alpha_{21}>\alpha_{22}, \alpha_{11}-\alpha_{12}\leq \alpha_{21}-\alpha_{22}$.}

We wish to achieve the sum-DoF value of $d_1 + d_2 =\min(\alpha_{11}+(\alpha_{21}-\alpha_{11}+\beta_1)^+,\alpha_{21}+\alpha_{12}-\alpha_{22}+\beta_2)$. Since $ \alpha_{11}-\alpha_{12}\leq\alpha_{21}-\alpha_{22}$, the first antenna can transmit $\alpha_{11}-\alpha_{12}$ DoF using its highest power levels and it will be decoded at the receivers without any interference from the second antenna. So, decreasing both $\alpha_{11},\alpha_{21}$ by $\alpha_{11}-\alpha_{12}$, we have a new channel with channel coefficients $\alpha'_{11}=\alpha_{12},\alpha'_{12}=\alpha_{12},\alpha'_{21}=\alpha_{21}-\alpha_{11}+\alpha_{12},\alpha'_{22}=\alpha_{22}$, where we need to achieve  the sum-DoF value of $d_1 + d_2 =\min(\alpha'_{11}+(\alpha'_{21}-\alpha'_{11}+\beta_1)^+,\alpha'_{21}+\alpha'_{11}-\alpha'_{22}+\beta_2)$ through the tuple $d_1 = \alpha'_{11}, d_2 =m$, where $m=\min((\alpha'_{21}-\alpha'_{11}+\beta_1)^+,\alpha'_{21}-\alpha'_{22}+\beta_2)$. Note that the case $m=0$ is obviously achievable, so lets consider the case where $m>0$. To achieve $\alpha'_{11}+m$ DoF, similar to first case, let us split User 1's message
as $W_1 = (W_c,W_{1z},W_{1p})$ and User 2's message as $W_{2} = W_{2z}$,
where $W_{1z},W_{1p}$ act as private sub-messages to be decoded
only by user 1, $W_{2z}$ acts as a private sub-message to be
decoded only by User 2, while $W_c$ acts as a common submessage
that can be decoded by both users. $W_c$, $W_{1z}$, $W_{2z}$ and $W_{1p}$
carry $\alpha'_{21}-m,m,m,\alpha'_{11}-\alpha'_{21}$ DoF respectively. Messages $W_c,W_{1z},W_{2z},W_{1p}$ are encoded into
independent Gaussian codebooks $X_c,X_{1z},X_{2z},X_{1p}$, with unit powers, producing the transmitted symbols as follows.
\begin{align}
\left[\begin{array}{c}
X_1\\
X_2
\end{array}
\right]=&~~c_o{\bf V}_cX_c+c_o\sqrt{P^{-\alpha'_{21}}}{\bf V}_{1p}X_{1p}+c_o{\bf V}_{1z} X_{1z}+c_o\sqrt{P^{m-\alpha'_{21}}}{\bf V}_{2z} X_{2z}
\end{align}
Here ${\bf V}_{c}$, ${\bf V}_{1p}$, ${\bf V}_{2p}$ are vectors as follows
\begin{eqnarray}
{\bf V}_{c}&=&\left[\begin{array}{c}
1\\
0
\end{array}
\right]\\
{\bf V}_{1p}&=&\left[\begin{array}{c}
1\\
0
\end{array}
\right]\\
{\bf V}_{1z}&=&\left[\begin{array}{c}
\hat{G}_{22}\sqrt{P^{m+\alpha'_{22}-2\alpha'_{21}}}\\
-\hat{G}_{21}\sqrt{P^{m-\alpha'_{21}}}
\end{array}
\right]\\
{\bf V}_{2z}&=&\left[\begin{array}{c}
\hat{G}_{12}\\
-\hat{G}_{11}
\end{array}
\right]
\end{eqnarray}
Thus, ${\bf V}_{1z}$ is a unit vector orthogonal to the estimated channel vector of User $2$, and ${\bf V}_{2z}$ is a unit vector orthogonal to the estimated channel vector of User 1. The private messages carried by symbols $X_{1z},X_{2z}$ are  zero-forced to the estimated channels of the undesired users, whereas the common message is  heard by both users. $c_o$ is a scaling factor, $O(1)$ in $P$, chosen to ensure that the transmit power constraint is satisfied. 
The  signal seen at Receiver 1 is,
\begin{eqnarray}
{Y}_1&=&\left[\begin{array}{cc}\bar{P}^{\alpha'_{11}}\hat{G}_{11}&\bar{P}^{\alpha'_{12}}\hat{G}_{12}\end{array}\right]\left[\begin{array}{c}
X_1\\
X_2
\end{array}
\right]+
\left[\begin{array}{cc}\bar{P}^{\alpha'_{11}-\beta_1}\tilde{G}_{11}&\bar{P}^{\alpha'_{12}-\beta_1}\tilde{G}_{12}\end{array}\right]\left[\begin{array}{c}
X_1\\
X_2
\end{array}
\right]+Z_1\nonumber\\
&=&\bar{P}^{\alpha'_{11}}c_1 X_c+\bar{P}^{\alpha'_{11}-\alpha'_{21}}c_2X_{1p}+\bar{P}^{\alpha'_{11}+m-\alpha'_{21}}c_3X_{1z}+\bar{P}^{m-\alpha'_{21}-\beta_1+\alpha'_{11}}c_4X_{2z}+Z_1\nonumber
\end{eqnarray}
where the $c_i$ are non-zero and bounded, i.e., $O(1)$  functions of $P$. Note that, $m-\alpha'_{21}-\beta_1+\alpha'_{11}\leq 0$.  Similar to the first case, with the similar approach and similar SINR values, Receiver 1 can decode $X_c,X_{1z},X_{1p}$ successfully.  The  signal seen at Receiver 2 is,
\begin{eqnarray}
{Y}_2&=&\left[\begin{array}{cc}\bar{P}^{\alpha'_{21}}\hat{G}_{21}&\bar{P}^{\alpha'_{22}}\hat{G}_{22}\end{array}\right]\left[\begin{array}{c}
X_1\\
X_2
\end{array}
\right]+
\left[\begin{array}{cc}\bar{P}^{\alpha'_{21}-\beta_2}\tilde{G}_{21}&\bar{P}^{\alpha'_{22}-\beta_2}\tilde{G}_{22}\end{array}\right]\left[\begin{array}{c}
X_1\\
X_2
\end{array}
\right]+Z_2\nonumber\\
&=&\bar{P}^{\alpha'_{21}}e_1 X_c+e_2X_{1p}+\bar{P}^{m-\alpha_{21}+\alpha_{22}-\beta_2}e_3X_{1z}+\bar{P}^{m}e_4X_{2z}+Z_1\nonumber
\end{eqnarray}
where the $e_i$ are non-zero and bounded, i.e., $O(1)$  functions of $P$. Note that, $m-\alpha_{21}+\alpha_{22}-\beta_2\leq 0$.  Similar to the first case, with similar SINR values, Receiver 2 can decode $X_c,X_{2z}$ successfully.

\item{$\alpha_{21}\leq\alpha_{22}$.}\\

We wish to achieve the sum-DoF value of $d_1 + d_2 =\min(\alpha_{11}+(\alpha_{22}-\alpha_{12}+\beta_1)^+,\alpha_{22}+\alpha_{11}-\alpha_{21}+\beta_2)$ through the tuple $d_1 = \alpha_{11}, d_2 =m$, where $m=\min((\alpha_{22}-\alpha_{12}+\beta_1)^+,\alpha_{22}-\alpha_{21}+\beta_2)$. The case $m=0$ is obviously achievable, so lets consider the case where $m>0$. Note that $m<\alpha_{22}$. To achieve $\alpha_{11}+m$ DoF, similar to first case, let us split User 1's message
as $W_1 = (W_c,W_{1z},W_{1p})$ and User 2's message as $W_{2} = W_{2z}$,
where $W_{1z},W_{1p}$ act as private sub-messages to be decoded
only by user 1, $W_{2z}$ acts as a private sub-message to be
decoded only by User 2, while $W_c$ acts as a common submessage
that can be decoded by both users. $W_c$, $W_{1z}$, $W_{2z}$ and $W_{1p}$
carry $\alpha_{22}-m,m,m$ and $\alpha_{11}-\alpha_{22}$ DoF, respectively. Messages $W_c,W_{1z},W_{2z},W_{1p}$ are encoded into
independent Gaussian codebooks $X_c,X_{1z},X_{2z},X_{1p}$, with unit powers, producing the transmitted symbols as follows.
\begin{align}
\left[\begin{array}{c}
X_1\\
X_2
\end{array}
\right]=&~~c_o{\bf V}_cX_c+c_o\sqrt{P^{-\alpha_{22}}}{\bf V}_{1p}X_{1p}+c_o{\bf V}_{1z} X_{1z}+c_o{\bf V}_{2z} X_{2z}
\end{align}
Here ${\bf V}_{c}$, ${\bf V}_{1p}$, ${\bf V}_{2p}$ are vectors as follows
\begin{eqnarray}
{\bf V}_{c}&=&\left[\begin{array}{c}
1\\
1
\end{array}
\right]\\
{\bf V}_{1p}&=&\left[\begin{array}{c}
1\\
1
\end{array}
\right]\\
{\bf V}_{1z}&=&\left[\begin{array}{c}
\hat{G}_{22}\sqrt{P^{m-\alpha_{22}}}\\
-\hat{G}_{21}\sqrt{P^{m+\alpha_{21}-2\alpha_{22}}}
\end{array}
\right]\\
{\bf V}_{2z}&=&\left[\begin{array}{c}
\hat{G}_{12}\sqrt{P^{m-\alpha_{11}+\alpha_{12}-\alpha_{22}}}\\
-\hat{G}_{11}\sqrt{P^{m-\alpha_{22}}}
\end{array}
\right]
\end{eqnarray}
So ${\bf V}_{1z}$ is a unit vector orthogonal to the estimated channel vector of User $2$, and ${\bf V}_{2z}$ is a unit vector orthogonal to the estimated channel vector of User 1. The private messages $X_{1z},X_{2z}$ are  zero-forced to the estimated channels of the undesired users, whereas the common message is heard by both users. $c_o$ is a scaling factor, $O(1)$ in $P$, chosen to ensure that the transmit power constraint is satisfied. 
The  signal seen at Receiver 1 is,
\begin{eqnarray}
{Y}_1&=&\left[\begin{array}{cc}\bar{P}^{\alpha_{11}}\hat{G}_{11}&\bar{P}^{\alpha_{12}}\hat{G}_{12}\end{array}\right]\left[\begin{array}{c}
X_1\\
X_2
\end{array}
\right]+
\left[\begin{array}{cc}\bar{P}^{\alpha_{11}-\beta_1}\tilde{G}_{11}&\bar{P}^{\alpha_{12}-\beta_1}\tilde{G}_{12}\end{array}\right]\left[\begin{array}{c}
X_1\\
X_2
\end{array}
\right]+Z_1\nonumber\\
&=&\bar{P}^{\alpha_{11}}c_1 X_c+\bar{P}^{\alpha_{11}-\alpha_{22}}c_2X_{1p}+\bar{P}^{\alpha_{11}+m-\alpha_{22}}c_3X_{1z}+\bar{P}^{m-\alpha_{22}-\beta_1+\alpha_{12}}c_4X_{2z}+Z_1\nonumber
\end{eqnarray}
where the $c_i$ are non-zero and bounded, i.e., $O(1)$  functions of $P$. Note that, $m-\alpha_{22}-\beta_1+\alpha_{12}\leq 0$. Similar to the first case, with SINR values of $P^{\alpha_{22}-m}$, $P^{m}$, and $P^{\alpha_{11}-\alpha_{22}}$, Receiver 1 can decode $X_c,X_{1z},X_{1p}$ respectively. The  signal seen at Receiver 2 is,
\begin{eqnarray}
{Y}_2&=&\left[\begin{array}{cc}\bar{P}^{\alpha_{21}}\hat{G}_{21}&\bar{P}^{\alpha_{22}}\hat{G}_{22}\end{array}\right]\left[\begin{array}{c}
X_1\\
X_2
\end{array}
\right]+
\left[\begin{array}{cc}\bar{P}^{\alpha_{21}-\beta_2}\tilde{G}_{21}&\bar{P}^{\alpha_{22}-\beta_2}\tilde{G}_{22}\end{array}\right]\left[\begin{array}{c}
X_1\\
X_2
\end{array}
\right]+Z_2\nonumber\\
&=&\bar{P}^{\alpha_{22}}e_1 X_c+e_2X_{1p}+\bar{P}^{m+\alpha_{21}-\alpha_{22}-\beta_2}e_3X_{1z}+\bar{P}^{m}e_4X_{2z}+Z_1\nonumber
\end{eqnarray}
where the $e_i$ are non-zero and bounded, i.e., $O(1)$  functions of $P$. Note that, $m+\alpha_{21}-\alpha_{22}-\beta_2\leq 0$. Similar to the first case, with the similar approach and similar SINR value, Receiver 2 can decode $X_c,X_{2z}$ successfully.

\end {enumerate}

\section{Proof of Theorem \ref{Theorem3}}
\subsection{Outer Bound}
The generalization of the proof to the $K$ user setting requires only a few extra steps for initial set up before the problem decomposes into the equivalent of what has been shown for the $K=2$ case. Here we describe the additional setup steps. 
%

Starting with the deterministic model, for the $k^{th}$ user we bound the rate as
\begin{eqnarray}
nR_k&\leq&I(W_k; \bar{Y}_k^{[n]}|G^n,W_{k+1}, W_{k+2}, \cdots, W_K)+o(n)\\
&\leq& H(\bar{Y}_k^{[n]}|G^n,W_{k+1}, \cdots, W_K)-H(\bar{Y}_k^{[n]}|G^n,W_k, W_{k+1}, \cdots, W_K)+o(n)
\end{eqnarray}
where  $G^n$ includes all channel realizations. Adding the rate bounds we obtain
\begin{eqnarray}
n\sum_{k=1}^KR_k&\leq&n\log(\bar{P})+\sum_{k=2}^K\left(H(\bar{Y}_{k-1}^{[n]}|G^n,W_{k},\cdots, W_K)-H(\bar{Y}_k^{[n]}|G^n,W_k, \cdots, W_K)\right)\nonumber
\end{eqnarray}
%
From this point on, the process of bounding the difference of entropy terms follows the proof of Theorem \ref{Theorem2}, so that we arrive at the bound
\begin{eqnarray}
n\sum_{k=1}^KR_k&\leq&n\log(\bar{P})+(1-\alpha+\beta)(K-1)n\log(\bar{P})
\end{eqnarray}
which bounds the total DoF by $1+(K-1)(1-\alpha+\beta)=(\alpha-\beta)+K(1-(\alpha-\beta))$.

\subsection {Achievability}
Let us prove that the sum-GDoF value of $\sum_{k=1}^Kd_k = 1+(K-1)(1-\alpha+\beta)$ is achievable through the $K$-tuple $d_1 = 1, d_k =1-\alpha+\beta$ for $k=2,3,\cdots, K$. To do this, let us similarly split User 1's message as $W_{1} = (W_c,W_{1p})$, where $W_{1p}$ acts as a private sub-message to be decoded only by User $1$, while $W_c$ acts as a common message that can be decoded by all users. The remaining messages $W_2, W_3, \cdots, W_K$ are all private, intended to be decoded only by their desired users. The common message $W_c$ carries $\alpha-\beta$ GDoF, whereas all private messages $W_{1p}, W_2, W_3, \cdots, W_K$
carry $1-\alpha+\beta$ GDoF each. Messages $W_c,W_{1p}, W_2, W_3, \cdots, W_K$ are encoded into 
unit power independent Gaussian codebooks $X_c,X_{1p}, X_{2p}, \cdots, X_{Kp}$,  respectively. 

The transmitted symbols are constructed as follows.
\begin{align}
\left[\begin{array}{c}
X_1\\
\vdots\\
X_K
\end{array}
\right]&=a\sqrt{1-P^{\beta-\alpha}}{\bf V}_cX_c
+a\sqrt{P^{\beta-\alpha}}\sum_{k=1}^K{\bf V}'_{kp}X_{kp}
\end{align}

Let $\hat{\bf G}$ be the $K\times K$ matrix whose $(k,l)^{th}$ term is defined as 
\begin{eqnarray}
\hat{\bf G}(k,l)&=&\left\{
\begin{array}{ll}
\hat{G}_{k,k},& k=l\\
\sqrt{P^{\alpha-1}}\hat{G}_{k,l},&k\neq l
\end{array}
\right.
\end{eqnarray}
The ${\bf V}'_{kp}$ are unit vectors chosen so that
\begin{align}
\hat{\bf G}
\left[\begin{array}{lll}{\bf V}'_{1p}&\cdots&{\bf V}'_{Kp}\end{array}\right]
\end{align}
is a diagonal matrix. In other words, the $k^{th}$ private message is sent in a direction orthogonal to the estimated channel vector of every user except the $k^{th}$ user. As before, ${\bf V}_c$ is a generic vector and $a$ is a scaling factor that is $O(1)$ in $P$, chosen to ensure that the transmit power constraint is satisfied.

 Let us take a closer look at the vectors ${\bf V}'_{kp}$. Consider, e.g., the product of the second row of $\hat{\bf G}$ and ${\bf V}'_{1p}$, scaled by $\sqrt{P^{1-\alpha}}$,
\begin{eqnarray}
\hat{G}_{2,1}{\bf V}'_{1p}(1) + \sqrt{P^{1-\alpha}}\hat{G}_{2,2}{\bf V}'_{1p}(2)+\hat{G}_{2,3}{\bf V}'_{1p}(3)+\cdots+\hat{G}_{2,K}{\bf V}'_{1p}(K)&=&0.
\end{eqnarray}
which implies that ${\bf V}'_{1p}(2)$ cannot be more than $O(\sqrt{P^{\alpha-1}})$ in $P$. Similarly, considering the product of the $m^{th}$ row of $\hat{\bf G}$ and ${\bf V}'_{1p}$, scaled by $\sqrt{P^{1-\alpha}}$, we note that ${\bf V}'_{1p}(m)$ cannot be more than $O(\sqrt{P^{\alpha-1}})$ in $P$, for $m\neq 1$. Proceeding similarly for the vector ${\bf V}'_{kp}$, considering the product of the $m^{th}$ row of $\hat{\bf G}$ and ${\bf V}'_{kp}$, scaled by $\sqrt{P^{1-\alpha}}$, we note that ${\bf V}'_{kp}(m)$ cannot be more than $O(\sqrt{P^{\alpha-1}})$ in $P$, for $m\neq k$. With this observation, we can define vectors ${\bf V}_{kp}$ whose elements are $O(1)$, such that 
\begin{align}
{\bf V}'_{kp}&={\bf M}_k{\bf V}_{kp}, ~~~~\forall k\in[K]
\end{align}
and ${\bf M}_k$ is a $K\times K$ diagonal matrix  with a $1$ as the $(k,k)^{th}$ element, all of whose remaining diagonal terms are equal to $\sqrt{P^{\alpha-1}}$.  

The  signal seen at Receiver 1 is,
\begin{eqnarray}
{Y}_1&=&\sqrt{P}\left[\begin{array}{lll}\hat{G}_{11}&\cdots&\hat{G}_{1K}\end{array}\right]{\bf M}_1\left[\begin{array}{c}
X_1\\
\vdots\\
X_K
\end{array}
\right]+\sqrt{P^{1-\beta}}
\left[\begin{array}{lll}\tilde{G}_{11}&\cdots&\tilde{G}_{12}\end{array}\right]{\bf M}_1\left[\begin{array}{c}
X_1\\
\vdots\\
X_K
\end{array}
\right]+Z_1~~~~~\\
&=&\sqrt{P}a_o X_c+\sqrt{P^{1+\beta-\alpha}}a_1X_{1p}+\sum_{k=2}^Ka_kX_{kp}+Z_1
\end{eqnarray}
where the $a_k$ are  $O(1)$  in $P$.

User 1 first decodes $X_c$ while treating all other signals as white noise. This is possible because  $X_c$ is received with power $\sim P$, the effective noise has power $\sim P^{1+\beta-\alpha}$, and $X_c$ carries $1-(1+\beta-\alpha)=\alpha-\beta$ GDoF. After decoding $X_c$, the receiver subtracts its contribution from its received signal and then proceeds to decode $X_{1p}$ while treating remaining signals as noise. Since $X_{1p}$ is received with power $\sim P^{1+\beta-\alpha}$, the remaining signals and noise are received with only  $O(1)$ power, and $X_{1p}$ carries $1+\beta-\alpha$ DoF, this decoding is successful as well. Thus, User 1 achieves $\alpha-\beta+1-\beta+\alpha=1$ GDoF. All other users proceed similarly to achieve $1-\beta+\alpha$ DoF, so that the total GDoF achieved equal $1+(K-1)(1-\beta+\alpha)$.

\section{Conclusion}
Because of the coarse and asymptotic character of DoF and GDoF metrics, even small gaps in our understanding of these coarse approximations can hide the most consequential ideas. Numerous discoveries around interference alignment emerged from efforts to find new achievable schemes to bridge the gap between the best inner and outer bounds. Following in the same spirit, this work bridges the extremes of known DoF results between perfect and finite precision CSIT. In the process, it expands our understanding of a relatively new idea -- the aligned image sets (AIS) approach. Interference alignment and AIS can be seen as two sides of the same coin. In the pursuit of DoF and GDoF characterizations, just as interference alignment enables powerful achievable schemes to close the gap from below, the AIS approach enables powerful outer bounds to close the gap from above. Whether these ideas are enough to close the GDoF gaps for all channels and regimes of interest, if so then what new insights emerge from the new GDoF characterizations, and if not, then what new ideas hide in the remaining gaps, are exciting questions for the future.

\bibliographystyle{IEEEtran}
\bibliography{Thesis}
\end{document}